\documentstyle[aps,prl,preprint]{revtex}
\tightenlines
\def\be{\begin{equation}}
\def\ee{\end{equation}}
\def\beq{\begin{eqnarray}}
\def\eeq{\end{eqnarray}}

\begin{document}
\draft
\preprint{gr-qc/0004015}
\title{Lorenz Gauge in Quantum Cosmology}
\author{Giampiero Esposito} 
\address{INFN, Sezione di Napoli, Complesso
Universitario di Monte S. Angelo, Via Cintia, Edificio N,
80126 Napoli, Italy}
\author{Giuseppe Pollifrone}
\address{Astronomy Unit, School of Mathematical Sciences,
Queen Mary \& Westfield College, London~E1~4NS, England}
\date{\today}
\maketitle
\begin{abstract}
In a path-integral approach to quantum cosmology, the Lorenz
gauge-averaging term is studied for Euclidean Maxwell theory
on a portion of flat four-space bounded by two 
concentric three-spheres, 
but with arbitrary values of the gauge parameter.
The resulting set of eigenvalue equations for normal and 
longitudinal modes of the electromagnetic potential cannot be
decoupled, and is here studied with a Green-function method.
This means that an equivalent equation for longitudinal modes is
obtained which has integro-differential nature, after inverting a
differential operator in the original coupled system. A 
complete calculational scheme is therefore obtained for the
one-loop semiclassical evaluation of the wave function of the
universe in the presence of gauge fields. This might also lead
to a better understanding of how gauge independence is actually
achieved on manifolds with boundary, whose consideration cannot
be avoided in a quantum theory of the universe.
\end{abstract}
\newpage
Over the last decade, much work in one-loop quantum cosmology has
been devoted to the analysis of the wave function of the universe
in the case of gauge fields on the Euclidean four-ball, or on a
portion of flat Euclidean four-space bounded by two concentric
three-spheres of radii $\tau_{-}$ and $\tau_{+}$. We are here 
interested in the electromagnetic case, since a proper understanding
of this model may already cast new light on quantum cosmology 
and quantum field theory. The normal and tangential components
of the electromagnetic potential are then expanded on a family of
concentric three-spheres in the form
\begin{equation}
A_{0}(x,\tau)=\sum_{n=1}^{\infty}R_{n}(\tau)Q^{(n)}(x) \; ,
\label{1}
\end{equation}
\begin{equation}
A_{k}(x,\tau)=\sum_{n=2}^{\infty}\Bigr[f_{n}(\tau)S_{k}^{(n)}(x)
+g_{n}(\tau)P_{k}^{(n)}(x)\Bigr]
\; \; \; \; {\rm for} \; {\rm all} \; k=1,2,3 \; ,
\label{2}
\end{equation}
where $Q^{(n)}(x),S_{k}^{(n)}(x),P_{k}^{(n)}(x)$ are scalar,
transverse and longitudinal vector harmonics on $S^{3}$, respectively.
Gaussian averages over gauge functionals are then performed according
to the Faddeev-Popov scheme, so that the part of the full Euclidean
action involving the potential $A_{\mu}$ reads [1]
$$
\int_{M}\left[
{1\over 4}F_{\mu \nu}F^{\mu \nu}
+{{\Bigr[\Phi(A)\Bigr]}^{2}\over 2\alpha}\right]
\; \sqrt{{\rm det} \; g} \; d^{4}x  \; ,
$$
where $F_{\mu \nu} \equiv
\partial_{\mu}A_{\nu}-\partial_{\nu}A_{\mu}$ denotes the
electromagnetic-field tensor, $g$ is the background four-metric, 
$\Phi$ is an arbitrary gauge-averaging functional
defined on the space of connection one-forms, and $\alpha$
is a dimensionless parameter. We are here interested in the
Lorenz choice for $\Phi$, i.e.
\begin{equation}
\Phi_{L} \equiv { }^{(4)}\nabla^{\mu}A_{\mu}=
{ }^{(4)}\nabla^{0}A_{0}
+ A_{0} \; {\rm Tr} \; K
+  { }^{(3)}\nabla^{i}A_{i} \; ,
\label{3}
\end{equation}
where $K$ is the extrinsic-curvature tensor of the three-sphere
boundary $S^{3}$, and ${ }^{(3)}\nabla$ is the induced connection
on $S^{3}$. The transverse modes $f_{n}$ are decoupled, whereas
longitudinal and normal modes turn out to obey the coupled
eigenvalue equations [1]
\begin{equation}
{\widehat A}_{n}g_{n}(\tau)+{\widehat B}_{n}R_{n}(\tau)=0 \; ,
\label{4}
\end{equation}
\begin{equation}
{\widehat C}_{n}g_{n}(\tau)+{\widehat D}_{n}R_{n}(\tau)=0 \; ,
\label{5}
\end{equation}
where the differential operators resulting from the choice (3) read
\begin{equation}
{\hat A}_{n} \equiv 
{d^{2}\over d\tau^{2}}
+{1\over \tau}{d\over d\tau}-{1\over \alpha}
{(n^{2}-1)\over \tau^{2}}+\lambda_{n} \; ,
\label{6}
\end{equation}
\begin{equation}
{\hat B}_{n} \equiv (n^{2}-1)\left[\left({1\over \alpha}-1
\right){d\over d\tau}+\left({3\over \alpha}-1 \right)
{1\over \tau}\right],
\label{7}
\end{equation}
\begin{equation}
{\hat C}_{n} \equiv  
\left(1-\frac{1}{\alpha}\right){1\over \tau^{2}}{d\over d\tau}
+\frac{2}{\alpha} {1\over \tau^{3}}\; ,
\label{8}
\end{equation}
\begin{equation}
{\hat D}_{n} \equiv  
{1\over \alpha}{d^{2}\over d\tau^{2}}
+{3\over \alpha}{1\over \tau}{d\over d\tau}
-\left({3\over \alpha}+n^{2}-1 \right)
{1\over \tau^{2}}+\lambda_{n} \; ,
\label{9}
\end{equation}
$\lambda_{n}$ being the eigenvalues [1].
For arbitrary values of $\alpha$ one cannot decouple Eqs. (4) and
(5) and map them into another system involving only differential operators.
One can however use with some profit integral equations and 
Green-kernel methods. For this purpose, we have to impose a suitable
set of boundary conditions. Since the tangential components of $A_{\mu}$
and the gauge-averaging functional should vanish at the boundary to
achieve gauge invariance of the whole set of boundary conditions [1,2],
we have
\begin{equation}
g_{n}(\tau_{+})=g_{n}(\tau_{-})=0,
\label{10}
\end{equation}
\begin{equation}
\left[{dR_{n}\over d\tau}+{3\over \tau}R_{n} \right]_{\tau=\tau_{+}}
=\left[{dR_{n}\over d\tau}+{3\over \tau}R_{n} 
\right]_{\tau=\tau_{-}}=0.
\label{11}
\end{equation}
These boundary conditions tell us how to proceed in order to solve the
system of coupled equations (4) and (5). Since we are aiming to invert
differential operators, it is clear that we have to consider operators
having well defined inverses once such boundary conditions are assigned.
Bearing this in mind, and denoting by ${\hat A}_{n}^{-1}$ the inverse
of ${\hat A}_{n}$, Eq. (4) leads to
\begin{equation}
g_{n}=-{\hat A}_{n}^{-1}{\hat B}_{n}R_{n},
\label{12}
\end{equation}
so that Eq. (5) implies
\begin{equation}
R_{n}={\cal P}_{n}R_{n},
\label{13}
\end{equation}
having defined
\begin{equation}
{\cal P}_{n} \equiv {\hat D}_{n}^{-1} \; {\hat C}_{n} \;
{\hat A}_{n}^{-1} \; {\hat B}_{n}.
\label{14}
\end{equation}
On denoting by $P_{n}(\tau,y)$ the kernel of ${\cal P}_{n}$, and
by $G_{n}(\tau,y)$ the Green kernel of ${\hat A}_{n}$, we 
therefore find (see below for the notation)
\begin{equation}
R_{n}(\tau)=\int_{\tau_{-}}^{\tau_{+}}P_{n}(\tau,y)R_{n}(y)dy,
\label{15}
\end{equation}
\begin{equation}
g_{n}(\tau)= -\int_{\tau_{-}}^{\tau_{+}}G_{n}(\tau,y)
({\hat B}_{n}R_{n})(y)dy.
\label{16}
\end{equation}
The general form of the kernels is obtained with the help of
standard techniques. For example, $G_{n}$ satisfies the differential
equation
\begin{equation}
{\hat A}_{n}G_{n}(\tau,y)=0 \; \; \; \; \; \; \forall \tau \not = y,
\label{17}
\end{equation}
as well as the continuity condition
\begin{equation}
\lim_{\tau \to y^{+}}G_{n}(\tau,y)
-\lim_{\tau \to y^{-}}G_{n}(\tau,y)=0,
\label{18}
\end{equation}
and the jump condition
\begin{equation}
\lim_{\tau \to y^{+}}{\partial \over \partial \tau}G_{n}(\tau,y)
-\lim_{\tau \to y^{-}}{\partial \over \partial \tau}
G_{n}(\tau,y)=1.
\label{19}
\end{equation}
Moreover, by virtue of (10), $G_{n}(\tau,y)$ obeys the boundary
conditions
\begin{equation}
G_{n}(\tau_{-},y)=0,
\label{20}
\end{equation}
\begin{equation}
G_{n}(\tau_{+},y)=0.
\label{21}
\end{equation}
On defining $\tau_{<} \equiv {\rm {min}}(\tau,y)$ and
$\tau_{>} \equiv {\rm {max}}(\tau,y)$, the general theory of
one-dimensional boundary-value problems [3] makes it therefore
possible to express the Green kernel $G_{n}(\tau,y)$ in the form
\begin{equation}
G_{n}(\tau,y)=C {u_{1}(M\tau_{<})u_{2}(M\tau_{>})\over 
W(u_{1},u_{2};\xi)},
\label{22}
\end{equation}
where $C$ is a constant, $\nu \equiv \sqrt{{n^{2}-1}\over \alpha}$, 
the parameter $M$ is the square root of $\lambda_{n}$,
$u_{1}$ and $u_{2}$ are linearly independent solutions of 
${\hat A}_{n}u=0$ vanishing at $\tau_{-}$ and $\tau_{+}$, respectively,
and having Wronskian $W(u_{1},u_{2};\xi)$. They can be chosen in the
form
\begin{equation}
u_{1}(\tau)=J_{\nu}(M\tau),
\label{23}
\end{equation}
\begin{equation}
u_{2}(\tau)={\widetilde A}J_{\nu}(M\tau)
+{\widetilde B}N_{\nu}(M\tau).
\label{24}
\end{equation}
The formula (22) should be inserted into Eq. (16) where $R_{n}$
(obtained from (15)) reads, more explicitly,
\begin{equation}
R_{n}(\tau)=\int_{\tau_{-}}^{\tau_{+}}dy \; \Gamma_{n}(\tau,y)
\int_{\tau_{-}}^{\tau_{+}}dz \Bigr({\hat C}_{n}G_{n}(y,z)\Bigr)
({\hat B}_{n}R_{n})(z),
\label{25}
\end{equation}
having denoted by $\Gamma_{n}(\tau,y)$ the Green kernel of ${\hat D}_{n}$.

We have therefore outlined a complete computational scheme for the 
evaluation of gauge modes in quantum cosmology when arbitrary values
of the gauge parameter $\alpha$ are considered. Interestingly, 
integral equations and Green-kernel methods seems to be unavoidable
technical steps if the Faddeev--Popov path integral for the wave 
functional is studied for all $\alpha$. In other words, even if local
boundary conditions are imposed with linear covariant gauges, the
contribution to functional determinants resulting from longitudinal
and normal modes involves, in general, a non-local analysis, as is
clear from the integral formulae (25) and (16). It now remains to be
seen whether the full $\zeta(0)$ value [1] is independent of $\alpha$.
Although the question remains unsettled, Eqs. (25) and (16) seem to
provide new tools for the solution of this longstanding problem in
one-loop quantum cosmology. It should also be stressed that 
gauge-invariant boundary conditions make it necessary to invert operators
of Bessel type (see (6) and (9)). In this sense, Bessel functions 
remain the fundamental tool for a flat-space analysis, even when gauge
modes remain coupled.


\begin{references}
\bibitem{[1]}
G. Esposito, A. Yu. Kamenshchik and G. Pollifrone, {\it Euclidean
Quantum Gravity on Manifolds with Boundary}, Fundamental Theories
of Physics, Vol. 85 (Kluwer, Dordrecht, 1997).
\bibitem{[2]}
I. G. Avramidi and G. Esposito, Commun. Math. Phys.
{\bf 200}, 495 (1999).
\bibitem{[3]}
I. Stakgold, {\it Green's Functions and Boundary Value Problems}
(Wiley, New York, 1979).

\end{references}
\end{document}